\documentclass[twocolumn]{aa}  

\usepackage{graphicx}
\usepackage{txfonts}
\begin{document}
\bibliographystyle{aa}

   \title{Bondi accretion for adiabatic flows onto a massive black hole with an accretion disk}

   \subtitle{The 1D problem}

   \author{
J. M. Ram{\'{\i}}rez-Vel{\'a}squez,
\inst{1,2}
L. Di G. Sigalotti,\inst{3} 
R. Gabbasov,\inst{4}
J. Klapp\inst{5,1}
and E. Contreras\inst{2}
          }

\institute{EPHYSLAB (Environmental Physics Laboratory), Facultad de Ciencias, Campus de Ourense, 
Universidad de Vigo, Ourense 32004, Spain.
\and
School of Physical Sciences and Nanotechnology, Yachay Tech University, 100119 Urcuqui, Ecuador.
\and
\'Area de F\'{\i}sica de Procesos Irreversibles, Departamento de Ciencias B\'asicas,
Universidad Aut\'onoma Metropolitana-Azcapotzalco (UAM-A), 
Av. San Pablo 180, 02200 Ciudad de M\'exico, Mexico.
\and
Instituto de Ciencias B\'asicas e Ingenier\'{\i}as, Universidad Aut\'onoma del Estado
de Hidalgo (UAEH), Ciudad Universitaria, Carretera
Pachuca-Tulancingo km. 4.5 S/N, Colonia Carboneras, Mineral de la Reforma, C.P. 42184,
Hidalgo, Mexico.
\and
Departamento de F\'{\i}sica, Instituto Nacional de Investigaciones Nucleares (ININ),
Carretera M\'exico-Toluca km. 36.5, La Marquesa,
52750 Ocoyoacac, Estado de M\'exico, Mexico.
}

   \date{Received September 15, 1996; accepted March 16, 1997}

 
  \abstract
   {{In this paper, we present the classical Bondi accretion theory for
the case of non-isothermal accretion processes onto a supermassive black hole (SMBH), 
including the effects of X-ray heating and the radiation force due to electron scattering 
and spectral lines.}{The radiation field is calculated by considering an optically thick,
geometrically thin, standard accretion disk as the emitter of UV photons and a spherical
central object as a source of X-ray emission. In the present analysis, the UV emission
from the accretion disk is assumed to have an angular dependence, while the X-ray/central
object radiation is assumed to be isotropic.} 
{This allows us to build streamlines in any
angular direction we need to.}
{The influence of both types of radiation
is evaluated for different flux fractions of the X-ray and UV emissions with and
without the effects of spectral line driving. We find that the radiation emitted near
the SMBH interacts with the infalling matter and modifies the accretion dynamics. In
the presence of line driving, a transition 
resembles
from pure type 1 \& 2 to type 5 solutions 
\citep[see Fig 2.1,][]{frank02a},
which takes place 
regardless of whether or not the UV emission 
dominates over the X-ray emission.}
{We compute the radiative factors at which this transition occurs, and discard type
5 solution from all our models.
Estimated values of the accretion radius and accretion rate in terms of the classical
Bondi values are also given. The results are useful for the construction of proper initial
conditions for time-dependent hydrodynamical simulations of accretion 
flows onto SMBH at
the centre of galaxies.}
}


   \titlerunning{Bondi accretion for adiabatic}
   \authorrunning{Ram{\'{\i}}rez-Vel{\'a}squez et al.}
   \maketitle
%

\section{Introduction}

The spherically symmetric Bondi solution \citep{Bondi1952} has become a reference model
for interpreting the observations of mass accretion onto supermassive black holes (SMBHs)
in the centre of galaxies. This is so because SMBHs at the centre of elliptical galaxies
are likely to accrete primarily from the surrounding hot, quasi-spherical interstellar
medium, suggesting that accretion rates can be simply estimated using Bondi accretion theory.
Studies on mass accretion were first considered in the framework of Newtonian gravity
\citep{bondi44a,Bondi1952,hoyle39a} and then they were generalized by \cite{michel72a} to 
curved spacetime. Quantum corrections to the general relativistic accretion have also been 
proposed \citep{yang15a,contreras18}.
In recent years, the accretion onto SMBHs has been the subject of intense research because
of its role in the co-evolution of SMBHs and their host galaxies. In particular, the number
of these studies down to the parsec scale 
in the centre of galaxies has significantly increased
with the advances in instrumental and computational capabilities
{\bf \citep[one example, is given by the][]{eht1}}. 
On the other hand, fully
analytical solutions on isothermal 
Bondi-like accretion including radiation pressure and the
gravitational potential of the host galaxy 
have started to appear recently 
\citep[hereafter KCP16,
CP17 and CZ18]{korol16a,ciotti17a,ciotti18a}. 
Moreover, recent detailed numerical calculations
on Bondi accretion \citep{ramirez18a}, 
using state-of-the-art consistent Smoothed Particle
Hydrodynamics (SPH) techniques 
\citep[]{gabbasov17a,sigalotti18a}, suggest that it would be
possible to push these studies further to 
cover the sub-parsec scales in active galactic nuclei
(AGNs), including the effects of radiation 
pressure due to spectral lines
\citep{ramirez16a,ramirez16b}.

Accretion onto compact objects is now 
recognized to be a major influencer on the environment
surrounding SMBHs in the centre of galaxies 
\citep[e.g.,][]{salpeter1964a,fabian1999a,barai2008a,germain2009a}.
In the study of the several properties governing the evolution of the system, i.e., 
Bondi quantities such as the accretion rate and the solution of the hydrodynamical equations,
a quite common assumption is to prescribe values of temperature, pressure and density at infinity 
and use them as the ``true" boundary conditions for the solution of the equations, even when the 
problem at hand needs sub-parsec resolution scales as,
for instance, in the case of AGNs. However, such resolution is never achieved in
observational and numerical studies 
\citep[]{silvia05}, introducing under- and
over-estimations of the physical quantities (CP17). 
The outflow phenomenon is believed to
play a major role in the feedback processes 
invoked by modern cosmological models (i.e.,
$\Lambda$-Cold Dark Matter) to explain the possible 
relationship between the SMBH and its
host galaxy 
\citep[e.g.,][]{magorrian1998a,gebhardt2000a} 
as well as in the self-regulating
growth of the SMBH. The problem of accretion onto a 
SMBH has been studied via
hydrodynamical simulations of galaxy evolution 
\citep[e.g.,][]{ciotti2001a,DiMatteo2005,li2007a,ostriker2010a,novak2011a}. 
For example, in numerical studies of galaxy formation, spatial
resolution permits resolving scales from the kpc to the pc, 
while the sub-parsec scales of
the Bondi radius are not resolved. 
This is why a prescribed sub-grid physics is often employed to
solve this lack of resolution. With sufficiently high X-ray luminosities, the falling material
will have the correct opacity, developing outflows that originate at sub-parsec scales
\citep[e.g.,][]{
ramirez2013a}. 
Therefore, calculations of the processes involving accretion onto SMBH have become of primary 
importance \citep[e.g.,][]{proga2000a,Proga2000b,proga2003a,pk2004,proga2007a,ostriker2010a}.
In order to provide a robust and reliable methodology for the production of initial conditions 
(ICs) for the numerical calculations of mass accretion onto SMBHs, we have combined the geometric
effects and assumptions employed by \cite{proga2007a} (hereafter P07) to compute the radiation
field from the disk and central object with the analytical solution procedure provided by KCP16
and CP17.

In particular, P07 reported axisymmetric, time-dependent hydrodynamical calculations of gas
flow under the influence of the gravity of black holes in quasars by taking into account X-ray
heating and the radiation force due to electron scattering and spectral lines. He found that for
a SMBH with a mass of $10^{8}M_{\odot}$ with an accretion luminosity of 0.6 times the Eddington
luminosity, the flow settles into a steady state and has two components: an equatorial inflow and
a bipolar inflow/outflow with the outflow leaving the system along the disk rotational axis and the
inflow being a realization of a Bondi-like accretion flow. To calculate the radiation field an
optically thin accretion disk was considered as a source of UV photons and a spherical central
object as a source of X-rays. In contrast, KCP16 generalized the classical Bondi accretion theory,
including the radiation pressure feedback due to electron scattering in the optically thin
approximation and the effects of a general gravitational potential due to a host galaxy.
In particular, they presented a full analytical discussion for the case of a Hernquist
galaxy model. An extension of this analytical solution was reported by CP17 in terms of
the Lambert-Euler $W$-function for isothermal accretion in Jaffe and Hernquist galaxies
with a central SMBH. They found that the flow structure is sensitive to the shape of the
mass profile of the host galaxy and that for the Jaffe and Hernquist galaxies the value
of the critical accretion parameter can be calculated analytically.

In this paper we derive radiative, non-isothermal ($\gamma\neq 1$), 
angular-dependent streamline
solutions for use as initial
conditions in numerical simulations of mass accretion flows onto massive compact objects. To do
so we introduce a radiation force term due to a non-isothermal extended source in the momentum
equation as in P07 and develop a semi-analytical solution for the non-isothermal accretion onto a
SMBH at the centre of galaxies using a procedure similar to that developed by KCP16 and CP17,
with no assumption about the Bernoulli equation, but radially integrating
the equations of motion.
The effects of the gravitational potential due to the host galaxy are ignored here, and they
will be left for a further analysis in this line. In Section 2, we introduce the mathematical
methodology and the fundamental equations, while Section 3 contains the analysis of the
results. 
Section 4 deals with the theoretical prediction of absorption lines.
Section 5 deals with a general analysis of the bias in the estimates of the Bondi radius
and mass accretion rate and discusses the importance of using the semi-analytical solution as true
initial conditions for numerical simulations of accretion flows. Finally, Section 6 contains
the relevant conclusions.

\section{Bondi accretion for a non-isothermal flow}

Under the assumption of spherical symmetry, 
the classical Bondi solution describes a purely
adiabatic accretion flow on a point mass, 
for the case of a gas at rest at infinity and
free of self-gravity. 
Detailed numerical calculations of the classical Bondi accretion flow
onto a stationary SMBH were performed by \cite{ramirez18a}, 
using a mathematically
consistent SPH method. 
In real situations, however, the accretion flow is affected by the
emission of radiation near the SMBH. The radiation emitted near 
the SMBH interacts with
the inflowing matter and modifies 
the accretion dynamics. The radiation effects can be
strong enough to stop the accretion and shut off the 
central object. Here we follow the
recipe of P07 to model the radiation 
field from the disk and central object, where a radiation
force term is added as a source in the momentum equation. 
In this formulation, the disk is
assumed to be flat, Keplerian, 
geometrically thin and optically thick. The structure and evolution
of a flow irradiated by a central compact object can be described 
by the
continuity and momentum equations
\begin{eqnarray}
\frac{d\rho }{dt}&=&-\rho\nabla\cdot {\bf v},\\
\frac{d{\bf v}}{dt}&=&-\frac{1}{\rho}\nabla p+\boldsymbol{\mathrm{g}}+{\bf F}^{\rm rad},
\end{eqnarray}
where $\rho$ is the mass density, 
${\bf v}=(v_{r},v_{\theta}=0,v_{\phi}=0)$ is the velocity
vector, $p$ is the gas pressure, 
$\boldsymbol{\mathrm{g}}$ is the gravitational acceleration due
to the central object, 
${\bf F}^{\rm rad}=(F_{r},F_{\theta}=0,F_{\phi}=0)$ is the total
radiation force per unit mass and $d/dt=\partial /\partial t +{\bf v}\cdot\nabla$.
Under the assumption that 
$v_{\theta}=v_{\phi}=0$, Eq. (2) reduces to a single equation
for the radial velocity component, 
which in spherical coordinates takes the form
\begin{eqnarray}
\frac{\partial v_{r}(r,\theta,\phi)}{\partial t}+
v_{r}(r,\theta,\phi)\frac{\partial v_{r}(r,\theta,\phi)}{\partial r}&=&
-\frac{1}{\rho}\frac{\partial p(r,\theta,\phi)}{\partial r}\nonumber\\
&-&\frac{\partial}{\partial r}\left(\frac{GM_{\rm BH}}{r}\right)\nonumber\\
&+&\frac{\partial}{\partial r}\left(\frac{C}{r}\right),
\end{eqnarray}
where $C$ is a constant whose value is 
given by the physical parameters
of the system,
while $\psi _{\rm grav}=(GM_{\rm BH})/r$ and 
$\psi _{\rm rad}=-C/r$,
are the gravitational and
radiation\footnote{We introduce this quantity for mathematical consistency.} 
potentials, respectively. The opposite signs in the potentials
indicate that the radiative and gravitational forces act in 
opposite directions. In the present
analysis we will study the 1D problem,
where
$v_{r}(r,\theta,\phi)=v_{r}(r,\theta_0)=v_{r}$,
for a fixed angle $\theta_0$,
and 
$p(r,\theta,\phi)=p(r,\theta_0)=p$. With these assumptions
Eq. (3) can be written in the more convenient form
\begin{equation}
\frac{\partial v_{r}}{\partial t}-p\frac{d}{dr}\left(\frac{1}{\rho}\right)+
\frac{d\mathcal{H}(r)}{d r}=0,
\end{equation}
where
\begin{equation}
\mathcal{H}(r)=\frac{p}{\rho}+\frac{1}{2}v_{r}^2-\psi _{\rm grav}-\psi _{\rm rad},
\end{equation}
is the so-called Bernoulli function $\mathcal{H}(r)$, which will be used below to
radially integrate equation (4).
In equations (4) and (5), the gas pressure is related to the mass 
density by the equation
\begin{equation}
p=\frac{k_{B}\rho T}{\mu m_{p}}=p_{\infty}\tilde{\rho}^{\gamma},
\end{equation}
where $k_{B}$ is the Boltzmann constant, $T$ is the gas temperature, $\mu$ is the mean
molecular weight, $m_{p}$ is the proton mass, $1<\gamma\leq 5/3$ is the polytropic index
and $\tilde\rho\equiv\rho /\rho _{\infty}$, where $p_{\infty}$ and $\rho _{\infty}$ are
the values of the pressure and density at infinity, respectively. The polytropic sound
speed is given by
\begin{equation}
c_{\rm s}^{2}=\frac{\gamma p}{\rho}.
\end{equation}

The radiative acceleration on the right-hand side of equation (3) is composed of two parts: 
one due to electron scattering, which was first studied for the isothermal case by KCP16, CP17 
and CZ18, and the other due to the radiation pressure produced by the spectral lines, which is here
designed using the same strategy developed by P07. The disk luminosity $L_{\rm disk}$ and the 
central object luminosity $L_{\star}$ are expressed in terms of the total accretion luminosity $L$. 
That is, $L_{\rm disk}=f_{\rm disk}L$ and $L_{\star}=f_{\star}L=(1-f_{\rm disk})L$. As in P07, the 
disk is assumed to emit only in the UV (i.e., $f_{\rm disk}=f_{\rm UV}$), while the central object 
is assumed to emit only X-rays (i.e., $f_{\star}=f_{\rm X}$). In this way, the central object
contributes to the radiation force due to electron scattering only, while the disk contributes to 
the radiation force due to both electron scattering and spectral lines. For simplicity, only the 
radiation force along the radial direction is considered, which for an optically thin gas is 
approximated by the relation 
\begin{equation}
F_{r}^{\rm rad}(r)|_{\theta=\theta_{0}}=\frac{\sigma _{T}L}{4\pi r^{2}cm_{p}}
\left[f_{\star}+2\cos\theta_{0}f_{\rm disk}(1+M(\tau))\right],
\end{equation}
where $c$ is the speed of light, $\sigma _{T}/m_{p}$ is the mass scattering coefficient for free
electrons, $\theta_{0}$ is some fixed and constant polar angle measured from the rotational
axis of the disk, $\tau$ is the optical depth 
and $M(\tau)$ is the so-called force multiplier
\citep{castor1975a}, 
which defines the numerical 
factor that parameterizes by how much spectral
lines increase the scattering coefficient. According to P07, we use the following prescriptions
for the force multiplier
\begin{equation}
M(\tau)=k\tau^{-\alpha}\left[\frac{(1+\tau _{\rm max})^{(1-\alpha)}-1}
{\tau _{\rm max}^{(1-\alpha)}}\right],
\end{equation}
where $k$ is a factor proportional to the total number of lines, $\alpha (=0.6)$ is the ratio
of optically thick to optically thin lines, 
$\tau _{\rm max}=\tau \eta _{\rm max}$ and
$\eta _{\rm max}$ is a parameter that determines the maximum value
$M_{\rm max}=k(1-\alpha)\eta _{\rm max}^{\alpha}$. The parameters $k$ and $\eta _{\rm max}$
are given by the following relations \citep{stevens90a}
\begin{equation}
k=0.03+0.385\exp\left(-1.4\xi ^{0.6}\right),
\end{equation}
and
\begin{equation}
	\ln\eta _{\rm max}=
\begin{cases}
	6.9\exp\left(0.16\xi ^{0.4}\right) ,& \quad {\rm if}\quad\ln\xi\leq 0.5,\\
	9.1\exp\left(-7.96\times 10^{-3}\xi\right), & \quad {\rm if}\quad\ln\xi >0.5,
\end{cases}
\end{equation}
where $\xi$ is the photoionization parameter. As in P07, the photoionization
parameter is calculated as $\xi =4\pi {\cal F}_{X}/n$, where ${\cal F}_{X}$ is the local
X-ray flux and $n$ is the number density of the gas given by $n=\rho/(m_{p}\mu)$, with
$\mu =1$. The local X-ray flux, corrected for the effects of optical depth, is defined
by
\begin{equation}
{\cal F}_{X}=\frac{L_{\star}}{4\pi r^{2}}\exp\left(-\tau _{X}\right),
\end{equation}
where $\tau _{X}$ is the X-ray optical depth, which can be estimated by the integral
\begin{equation}
\tau _{X}=\int _{0}^{r}\kappa _{X}\rho dr,
\end{equation}
between the central source ($r=0$) and a point $r$ in the accreting flow, where
$\kappa _{X}$ is the absorption coefficient. Here the attenuation of the X-rays is
calculated using $\kappa _{X}=0.4$ g$^{-1}$ cm$^{2}$ for all $\xi$.

Introducing the normalized quantities
\begin{equation}
x\equiv\frac{r}{r_{B}},
\quad\tilde{c_{\rm s}}\equiv\frac{c_{\rm s}}{c_{\infty}}=
\tilde{\rho }^{(\gamma -1)/2},
\quad\mathcal{M}\equiv\frac{{v}_r}{c_{\rm s}},
\end{equation}
where
\begin{equation}
r_{B}=\frac{GM_{\rm BH}}{c_{\infty}^{2}},
\end{equation}
is the Bondi radius, $c_{\infty}^{2}=\gamma p_{\infty}/\rho _{\infty}$, 
$G$ is the Newtonian
gravitational constant, $M_{\rm BH}$ is the mass of the
black hole and $\mathcal{M}$ is the
Mach number, and assuming a steady state motion (i.e., $\partial v_{r}/\partial t=0$),
Eqs. (4) and (5) can be combined to give
\begin{equation}
\int\frac{d}{dr}\left(\frac{\mathcal{M}^{2}\tilde{c_{\rm s}}^{2}}{2}+
\frac{\tilde{\rho}^{(\gamma -1)}}{\gamma -1}-\frac{1}{x}+
\frac{l_{\rm tot}^{\rm rad}|_{\theta=\theta _{0}}}{x}\right)dr=0,
\end{equation}
where $C|_{\theta =\theta _{0}}=GM_{\rm BH}l_{\rm tot}^{\rm rad}|_{\theta =\theta _{0}}$.
The above integral is easily evaluated as follows
\begin{eqnarray}
\int_{\mathcal{M}_{\infty}}^{\mathcal{M}}
d\left(\frac{\mathcal{M}^{2}\tilde{c_{\rm s}}^{2}}{2}\right)=
\frac{\mathcal{M}^{2}\tilde{c_{\rm s}}^{2}}{2}\\
\int_{\rho _{\infty}}^{\rho}
d\left[\frac{\tilde{\rho}^{(\gamma -1)}}{\gamma -1}\right]=
\frac{1}{\gamma -1}\left[\tilde{\rho}^{(\gamma -1)}-1\right]\\
\int_{r_{\infty}}^{r}
d\left(-\frac{1}{x}+\frac{l_{\rm tot}^{\rm rad}|_{\theta=\theta _{0}}}{x}\right)=
-\frac{1}{x}+\frac{l_{\rm tot}^{\rm rad}|_{\theta=\theta _{0}}}{x},
\end{eqnarray}
where $\mathcal{M}_{\infty}\to 0$ and $r_{\infty}\to\infty$.
Integration of 
equation (16) leads to:
\begin{equation}
\tilde{\rho}^{(\gamma -1)}\left(\frac{\mathcal{M}^{2}}{2}+\frac{1}{\gamma -1}\right)=
\frac{1}{x}-\frac{l_{\rm tot}^{\rm rad}|_{\theta=\theta _{0}}}{x}+\frac{1}{\gamma -1},
\end{equation}
for the steady-state, radial, non-isothermal ($\gamma >1$), gravitational
accretion, including the effects of radiation emission due to electron scattering and
spectral discrete lines with appropriate boundary conditions at infinity, where
\begin{equation}
l_{\rm tot}^{\rm rad}|_{\theta=\theta _{0}}=
l_{\rm Edd}^{\rm rad}f^{\rm rad}|_{\theta=\theta _{0}},
\end{equation}
$l_{\rm Edd}^{\rm rad}=L/L_{\rm Edd}$, $L_{\rm Edd}=4\pi cGM_{\rm BH}m_{p}/\sigma _{T}$ is
the Eddington luminosity, $\sigma _{T}=6.6524\times 10^{-25}$ cm$^{2}$ is the Thomson cross
section and $f^{\rm rad}|_{\theta=\theta _{0}}$ is the radiative force parameter given by
\begin{equation}
f^{\rm rad}|_{\theta=\theta_0}=f_{\star}+2\cos\theta _{0}f_{\rm disk}[1+M(\tau)].
\end{equation}
Although the force multiplier depends on the gas temperature through the parameter $k$, as
shown by equation (17) of P07 based on detailed photoionization calculations performed with
the XSTAR code (T. Kallman 2006, private communication), here we adopt the temperature-independent
relation given by their equation (12) and leave the corrections for the temperature effects
for future numerical work in this line (for instance, when using an energy 
equation accompanied by the net heating and cooling function developed 
by \cite{ramirez16a,ramirez16b}.{\bf For example, see also \cite{danne18a}.)}
{\bf 
The temperature correction for a fixed ionization parameter $\xi$ is explained in 
detail in P07, but here we provide a brief qualitative explanation. The dependence of the 
force multiplier on both $\xi$ and $T$ is an extremely complex problem, in part due to 
the amount and participation of the $\sim 10^4$ different atomic transitions given by 
neutral and ionized species from H to Fe. It is worth to explore in detail these 
relationships. However, we know that $r\sim 10$~pc is a good place to
start\footnote{{\bf For ultraluminous X-ray sources (ULXs) 
$L\sim 10^{46},~ n_H\sim 10^{(7-9)}$ and $\xi\sim (1-10)$, $r\sim (1-10){\rm pc}$ 
\citep{pinto2019a}. Physically, this distance seems to be a plausible location
for super-Eddington outflows to take place given the relationship between the 
hardness ratio and the ionization parameter as well as the role that radiation 
pressure could be playing in these objects.}}
such a study due to the amount of UV absorption transitions available for spectral 
line acceleration \citep{kashi13a,higg14a,ramirez16a,ramirez16b} 
and is suitable for the comparison we are proposing in this work. So we are computing the 
radiative factor for a $\xi_{\rm 10pc}=\xi(r={\rm 10~pc})$ without temperature corrections 
even though the hydrodynamical variables are taken to be temperature dependent. For
instance, the dependence on temperature and ionization can be taken into account using
the detailed \cite{ramirez16b}'s Tables.}
Equation (20) must be solved coupled to the continuity equation (1), which in terms of the
above assumptions and normalized parameters can be written as
\begin{equation}
x^2\mathcal{M}\tilde{\rho}^{(\gamma +1)/2}=\lambda,
\label{Mdot1}
\end{equation}
where
\begin{equation}
\lambda =\frac{\dot{M}_{B}}{(4\pi f_{\rm solid}) r_{B}^{2}\rho_{\infty}c_{\infty}},
\end{equation}
is the dimensionless accretion parameter that determines the accretion rate for given
$M_{\rm BH}$ and boundary conditions. In Eq. (24)
$4\pi f_{\rm solid}=\int\sin\theta d\theta d\phi$ is the solid angle covered by the streamline
at the polar angle $\theta _{0}$. 
When $f_{\rm solid}=1$ (full solid angle), we recover the
classical accretion parameter. 
However, for the case of a $\theta _{0}$-dependent force
$f_{\rm solid}<<1$. 
This dependence is important only for the final calculation of
$\dot{M}_{B}=\lambda (4\pi f_{\rm solid}) r_{B}^{2}\rho _{\infty}c_{\infty}$, which is
fixed once the value of $\theta _{0}$ is chosen.
Using Eq. (23) to eliminate $\tilde\rho$ 
from equation (20), the Bondi problem
reduces to solving the equation:
\begin{equation}
g(\mathcal{M})=\Lambda f(x), \quad {\rm with}\quad
\Lambda =\left[\chi _{\rm tot}^{\rm rad}
|_{\theta=\theta_0}^{2}\lambda _{\rm cr}\right]^{2(1-\gamma)/(\gamma +1)},
\label{gM1}
\end{equation}
where $\chi _{\rm tot}^{\rm rad}|_{\theta=\theta_0}=1-l_{\rm tot}^{\rm rad}|_{\theta=\theta_0}$,
$\lambda$ is set equal to the critical value of the accretion parameter, defined as
$\lambda =\chi _{\rm tot}^{\rm rad}|_{\theta=\theta _{0}}^{2}\lambda _{\rm cr}$, and
\begin{eqnarray}
g(\mathcal{M})&=&\mathcal{M}^{2(1-\gamma)/(\gamma +1)}\left(\frac{\mathcal{M}^{2}}{2}
+\frac{1}{\gamma -1}\right),\label{gM2}\\
f(x)&=&x^{4(\gamma -1)/(\gamma +1)}\left(\frac{\chi _{\rm tot}^{\rm rad}|_{\theta=\theta_0}}{x}
+\frac{1}{\gamma -1}\right),\label{gM3}\\
\lambda _{\rm cr}&=&\frac{1}{4}\left(\frac{2}{5-3\gamma}\right)^{(5-3\gamma)/[2(\gamma -1)]}\label{gM4}.
\end{eqnarray}
Although equation (25) together with the 
definitions (26)-(28) look the same as those derived
by KCP16 and CP17, there are some differences 
that are described in the following. On the
other hand, it is clear from equation (25) 
that all physical quantities can be expressed in terms
of the radial Mach number profile.
\begin{figure}
\begin{center}
\includegraphics[width=8.5cm]{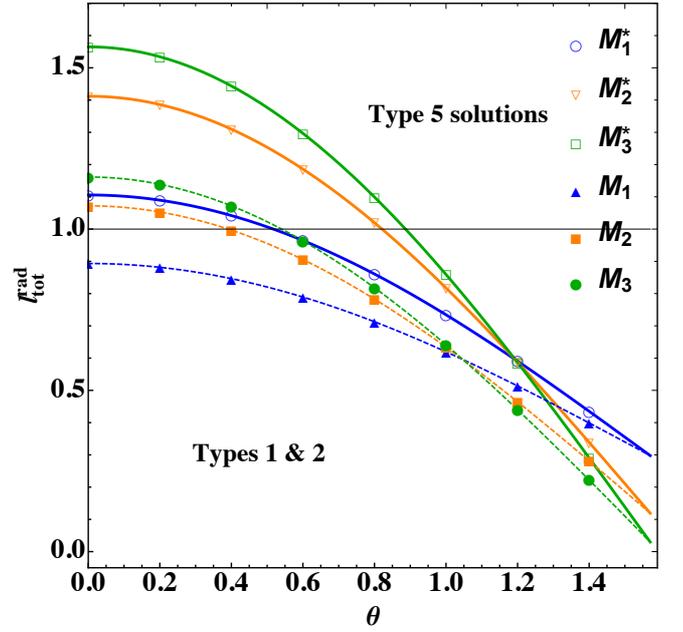}
\end{center}
\caption{Angular dependence of the radiative
force parameter for models $M_{1}$, $M_{2}$ and
$M_{3}$ (when $M(\tau)=0$; thin dashed lines) and 
models $M_{1}^{\star}$, $M_{2}^{\star}$ and
$M_{3}^{\star}$ (when $M(\tau)\neq 0$; thick lines). 
The horizontal line marks the
value $l_{\rm tot}^{\rm rad}=1$. 
Just below this value, the matter will undergo a pure
accretion process (type 1 solution), 
while above this value, 
it will suffer a transition to a type 5 outflow solution 
(which we discard). Each pair
of models ($M_{i}$, $M_{i}^{\star}$) uses the 
same boundary conditions.} 
\end{figure}

Compared to the previous analyses 
performed by KCP16 and CP17, the new feature of the present
model is the angular dependence of the UV emission 
of the accretion disk, while it is assumed
that the X-ray/central object radiation does not 
change with $\theta _{0}$.
For an incident angle of $\theta _{0}=0$, 
the particles experience maximum intensity.
As $\theta _{0}$ increases towards the equator, 
it follows from equation (22) that the
radiation flux from the accretion disk becomes 
weaker until it vanishes for $\theta _{0}=\pi /2$.
Consequently, the ratio between the X-ray and the UV 
flux increases with increasing $\theta _{0}$.
Thus, for large $\theta_0$ pure gravitational 
accretion (type 1 solution)\footnote{The type 2 transonic solution exists but 
the boundary conditions are imposed for an accretion flow 
(i.e., at infinity, [\citep{waters12a}]).} 
takes place until a critical angle
$\theta ^{\rm phase}$ develops for 
which radiation dominates over gravity and type 5 solutions result.
In Fig. 1 we plot this feature of the solutions for three 
different flux fractions of the X-ray
and UV emissions with and without the 
force multiplier. For instance, models
$M_{1}$, $M_{2}$ and $M_{3}$ have $f_{\star}=0.5$ 
and $f_{\rm disk}=0.5$, $f_{\star}=0.2$ and
$f_{\rm disk}=0.8$ and $f_{\star}=0.05$ and 
$f_{\rm disk}=0.95$, respectively, with $M(\tau)=0$
so that we can evaluate the influence of both 
types of radiation in detail. Models $M_{1}^{*}$,
$M_{2}^{*}$ and $M_{3}^{*}$ are identical to 
models $M_{1}$, $M_{2}$ and $M_{3}$, respectively,
but with $M(\tau)\neq 0$ and a radially varying 
ionization parameter given by
\begin{equation}
\xi =\frac{L}{n_{\rm H}r^{2}},
\end{equation}
where the total accretion luminosity 
is set to be $L=7.45\times 10^{45}$ erg s$^{-1}$, 
which is
appropriate for a SMBH of $10^{8}M_{\odot}$ 
at a distance $r=10$ pc, accreting at an efficiency
of 8\%. The gas density is $n_{\rm H}=10^{10}$ cm$^{-3}$ 
and the optical depth is set to $\tau=0.3$.
The physical parameters employed 
for all these models are listed in Table~\ref{tbl1}.
For all models we set $\theta _{0}=\theta_{\rm acc}^{\rm phase}$
\footnote{And for $M_1$ we set $\theta _{0}=\pi/4$.} 
and
$\gamma =1.1$. {\bf The exploration of different values of the polytropic
index $\gamma$ is a very complex topic. For instance, the ``mirror" problem,
Parker's winds, with $3/2<\gamma<5/3$, and their spherical properties do change
when angular momentum is added to the flow equations \citep{waters12a}.
Also $\gamma=3/2$ is the critical adiabatic index that separates the solution
space from their transonic behaviour. So, we plan to carefully look at $\gamma <3/2$
and $\gamma >3/2$ solutions, which lead to accelerating and decelerating Parker winds,
respectively}. 
The values of $f_{\star}$ and
$f_{\rm disk}$ listed in Table 1 are the same employed 
by P07 and are guided by the observational
results from \cite{zheng1997a} and \cite{laor1997a}.
\begin{table*}
 \centering
 \begin{minipage}{140mm}
  \caption{Parameters of the generalized Bondi accretion models.}
  \label{tbl1}
  \begin{tabular}{@{}ccccccccc@{}}
  \hline
{Run}&
Lines$^{(a)}$& 
{$f_{\rm \star}$}& 
{$f_{\rm disk}$ }& 
{$\gamma$}&
{$\theta^{\rm phase}$}&
{$\theta^{\rm phase}_{\rm acc}$}&
{$\theta^{\rm phase}_{\rm T5}$}&
{$x_{\rm crit}$}\\
& & & & & (rad) & (rad) & (rad) &\\
\hline
M$_{\rm 1}^*$~~~~~ & yes &   0.50 & 0.50 & 1.1 & ~$0.517465$ & 1.1 & 0.9 & 0.00856~~  \\
M$_{\rm 2}^*$~~~~~ & yes &   0.20& 0.80 & 1.1 & ~$0.821151$ & 1.1 & 0.9 & 0.03350~~  \\
M$_{\rm 3}^*$~~~~~ & yes &   0.05& 0.95 & 1.1 & $0.886684$ & 1.1 & 0.9 & 0.04559~~ \\
M$_{\rm 1}$~~~~~ & no &   0.50 & 0.50 & 1.1 & ~$na$ & 1.1 & 0.9 & 0.12149~~  \\
M$_{\rm 2}$~~~~~ & no &   0.20& 0.80 & 1.1 & ~$0.391333$ & 1.1 & 0.9& 0.00571~~  \\
M$_{\rm 3}$~~~~~ & no &   0.05& 0.95 & 1.1 & $0.540618$ & 1.1 & 0.9& 0.01332~~ \\
\hline
\end{tabular}\\
(a) $M(\tau)\neq 0$ (yes) and $M(\tau)=0$ (no).\\
The angles $\theta^{\rm phase}_{\rm acc}$ and 
$\theta^{\rm phase}_{\rm T5}$ are given in 
terms of $\theta^{\rm phase}$, i.e.,
$\theta^{\rm phase}_{\rm acc}=1.1\theta^{\rm phase}$ and
$\theta^{\rm phase}_{\rm T5}=0.9\theta^{\rm phase}$.
\end{minipage}
\end{table*}

Models $M_{1}$, $M_{2}$ and $M_{3}$ 
represent non-isothermal accretion models with only the
effect of electron scattering as in KCP16, CP17 and CZ18. 
A look to Fig. 1 shows that for model
$M_{1}$, where the X-ray and UV emitters 
have each the same fraction of participation (50\%) in
the emission of radiation, there is no angle at which 
type 5 solutions are produced. In fact, model
$M_{1}$ represents a pure accretion 
process with a nearly cancellation of the force of gravity
close to $\theta _{0}=0$ and a radiative force parameter,
$l_{\rm tot}^{\rm rad}|_{\theta =\theta _{0}}$, always below unity. 
In contrast, model $M_{1}^{*}$
with $M(\tau)\neq 0$ and $\theta ^{\rm phase}\approx 0.51$ 
undergoes a transition from pure accretion
to type 5 solution
with $l_{\rm tot}^{\rm rad}|_{\theta= \theta _{0}}\approx 1.1$ at
$\theta _{0}\approx 0$. 
Similarly, models $M_{2}$ and 
$M_{3}$ have values of $\theta ^{\rm phase}$
for which there will be non physical solutions around 
$\approx 0.4-0.5$, 
while models $M_{2}^{*}$ and $M_{3}^{*}$ develop
type 5 solutions for values of $\approx 0.82-0.90$.
In what follows we shall explore in more detail these numerical
solutions separately.

\section{Analysis of the results}

\begin{figure}
\begin{center}
\includegraphics[width=8.5cm]{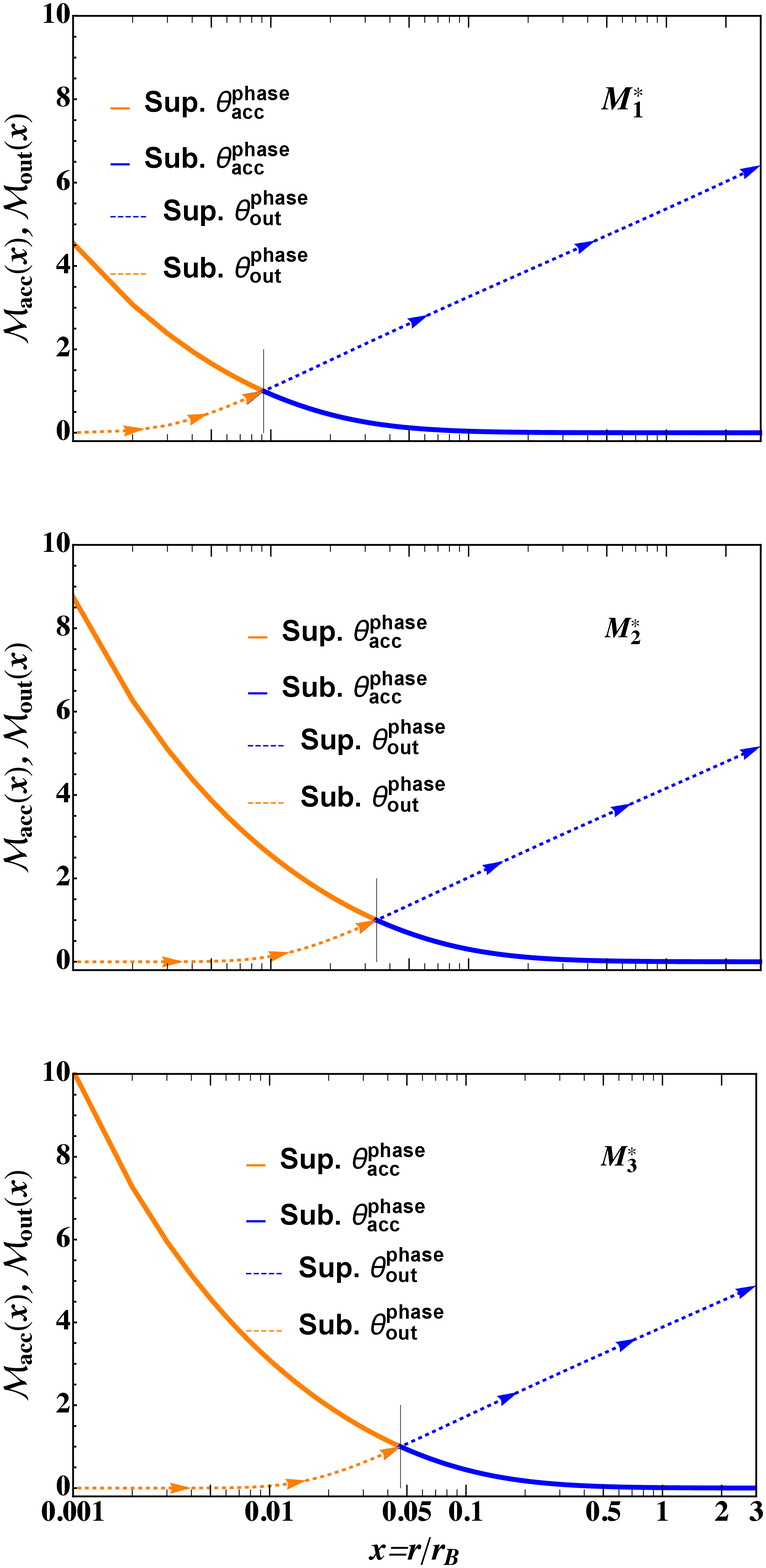}
\end{center}
\caption{Mach number of the inflow 
(solid lines) and outflow (dotted lines) solutions of
equation (20) with $M(\tau)\neq 0$. 
The supersonic part of the inflow solutions are depicted
in orange, 
while the subsonic part is given in blue. 
In contrast, the supersonic outflow
solution is depicted in blue and the 
subsonic one in orange. The vertical lines in each
frame mark the position of the sonic radius. 
When the radiation dominates, the gas will
escape from the gravitational 
potential of the accretor. The upper, middle and bottom
frames correspond to models with $f_{\rm disk}=0.5, 0.8$ and 0.95, respectively (see Table
\ref{tbl1}).}
\end{figure}
\begin{figure}
\begin{center}
\includegraphics[width=8.5cm]{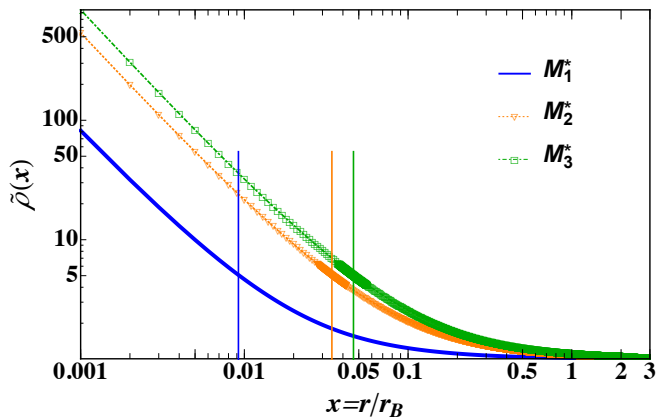}
\end{center}
\caption{Density profiles of the 
radiative non-isothermal models with $M(\tau)\neq 0$. The
vertical lines mark the position of the sonic radius 
for each model. As the UV emission
becomes stronger, the accreting gas 
becomes supersonic at larger radial distances from
the SMBH, while the density becomes higher for $x\lesssim 0.5$.}
\end{figure}

Figure 2 shows the radial Mach number 
profiles of the inflow (solid lines) and type 2 
(dotted lines) solutions 
(which we are including into the 
discussion and which we will call outflow
from now on)
of equation (25) for 
models $M_{1}^{*}$, $M_{2}^{*}$ and $M_{3}^{*}$. The
supersonic inflow ($\mathcal{M}>1$) solution is 
depicted in orange, while the subsonic
($\mathcal{M}<1$) solution is given by 
the blue lines. For model $M_{1}^{*}$, X-ray heating
is the strongest (i.e., $f_{\star}=f_{\rm disk}=0.5$). 
In this case, the solution predicts
a transition from pure accretion 
(i.e., $\theta _{0}>\theta ^{\rm phase}$)
to a 
type 5 flow
with a critical radius $x_{\rm crit}\approx 0.0090$, 
where the supersonic and subsonic flows
match to produce a full transonic flow. 
For model $M_{2}^{*}$, X-ray heating by the central
object ($f_{\star}=0.2$) is smaller than the UV 
emission from the accretion disk
($f_{\rm disk}=0.8$) and we observe 
two important differences compared to model $M_{1}^{*}$.
First, the critical point occurs at a larger 
radius ($x_{\rm crit}\approx 0.034$) and so the
flow becomes supersonic at a distance from the SMBH 
about 4 times larger than for model
$M_{1}^{*}$. Second, at an inner radius 
of $\approx 0.001$, the inflow for model $M_{2}^{*}$
reaches $\mathcal{M}\approx 8.1$ against 
$\approx 4.1$ for model $M_{1}^{*}$.

In Fig. 2, the supersonic outflow solution 
(type 2) is depicted by the
dotted blue lines, while the subsonic 
solutions are given by the dotted orange lines. 
Notice that
for the $M_{2}^{*}$ model, 
the winds are developed farther away 
from the centre, i.e., around
$x\approx 0.034$ compared to model $M_{1}^{*}$, 
where the thermal outflows are produced closer
to the central object at $x\approx 0.01$. 
Therefore, at larger distances from the SMBH, i.e., at
$x\approx 3$, faster winds are developed for 
model $M_{1}^{*}$ ($\mathcal{M}(x=3)\approx 6.2$)
compared to model $M_{2}^{*}$ for 
which $\mathcal{M}(x=3)\approx 4.4$.
In all plots, the 
flows escaping supersonically from
the gravitational potential of the SMBH are indicated 
by the arrows. When only 5\% of the
heating from the central object contributes 
to the radiation factor, as in model $M_{3}^{*}$,
$x_{\rm crit}\approx 0.046$ and mass accretion 
proceeds supersonically below this radius with
$\mathcal{M}\approx 10$ at $x=0.001$, while 
supersonic outflows are
produced for $x\gtrsim 0.046$. 
The numerical simulations of P07 for $f_{\star}=f_{\rm disk}=0.5$
(his run case A) shows that a strong X-ray heating can accelerate 
the outflow to maximum
velocities of 700 km s$^{-1}$ 
with the outflow collimation by the infall increasing with
increasing radius. When $f_{\star}=0.2$ and $f_{\rm disk}=0.8$, 
line driving is seen to
accelerate the outflow to higher velocities 
(up to 4000 km s$^{-1}$), while the outflow
collimation becomes very strong for 
$f_{\star}=0.05$ and $f_{\rm disk}=0.95$, i.e., when
the X-ray heating is the smallest, which 
corresponds to our model $M_{3}^{*}$. In this case,
the gas outflow is confined within a very 
narrow channel along the equatorial axis of the
accretion disk, while most of the 
computational volume is occupied by the inflow.

The density profiles for models $M_{1}^{*}$, 
$M_{2}^{*}$ and $M_{3}^{*}$ are shown in Fig. 3.
As the UV emission intensity from the accretion 
disk becomes stronger, the gas density
close to the SMBH becomes higher. This is 
counter-intuitive because we would have expected the
radiation to push the gas away from the centre, 
resulting in decreasing density close to
the SMBH. This is in fact the case. 
As the radiation intensity increases, it pushes the
critical radius farther away from the SMBH 
and consequently the gas becomes supersonic at
larger radial distances. As the supersonic 
flow occupies a larger central volume, the density
close to the SMBH increases by factors as 
large as $\approx 10$ when going from
$f_{\rm disk}=0.5$ to $f_{\rm disk}=0.95$.
\begin{figure}
\begin{center}
\includegraphics[width=8.5cm]{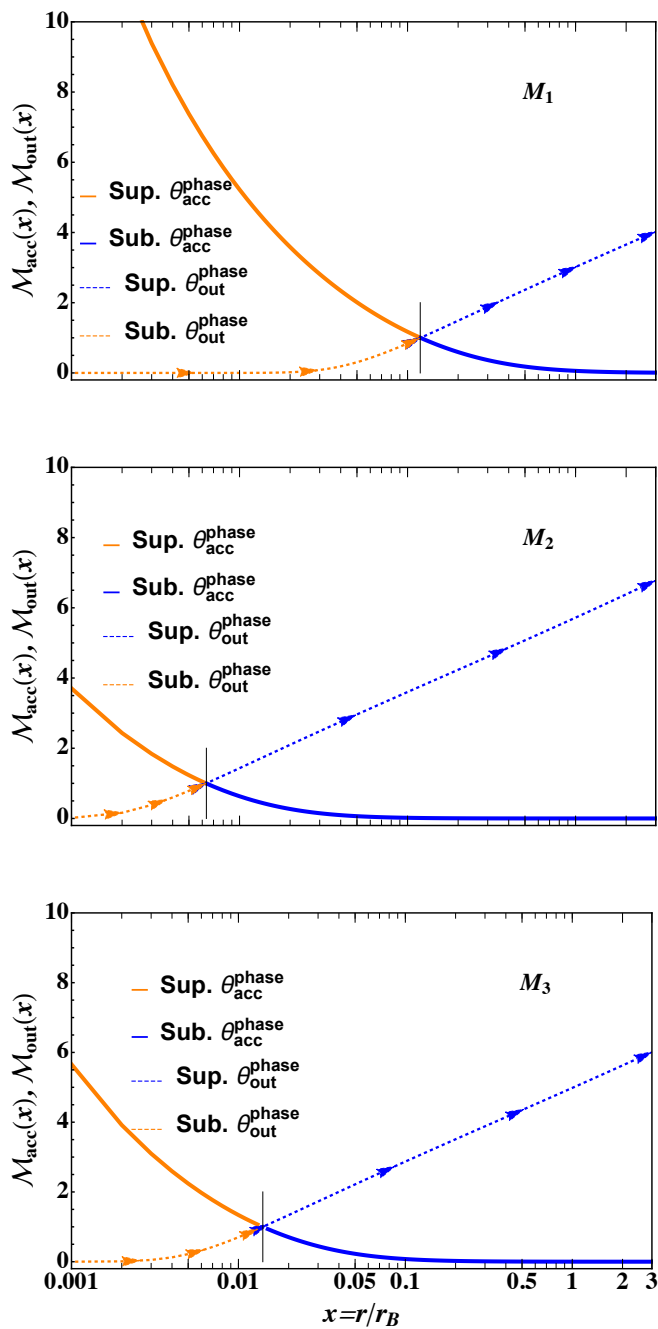}
\end{center}
\caption{Mach number of the inflow 
(solid lines) and outflow (dotted lines) solutions of
equation (20) with $M(\tau)= 0$. The supersonic part of 
the inflow solutions are depicted
in orange, while the subsonic part 
is given in blue. In contrast, the supersonic outflow
solution is depicted in blue and the subsonic one in orange. 
The vertical lines in each
frame mark the position of the sonic radius. 
When the radiation dominates, the gas will
escape from the gravitational potential of 
the accretor. With no force multiplier, no
winds are developed for $f_{\star}=f_{\rm disk}=0.5$ 
(model $M_{1}$).}
\end{figure}

The Mach number profiles for 
models $M_{1}$, $M_{2}$ and $M_{3}$ with no force multiplier
are shown in Fig. 4. In the absence of line 
driving and strong X-ray heating (model $M_{1}$),
no type 5 flows are developed regardless of 
the angle of incidence. In this case, the critical point
occurs at $x\approx 0.12$ and close to the SMBH at 
$x=0.001$ the Mach number reaches values as
high as $\approx 250$. As the UV emission 
intensity increases over the X-ray intensity (models
$M_{2}$ and $M_{3}$), the critical point and 
the outflows occur at smaller radii from the SMBH
compared to models $M_{2}^{*}$ and $M_{3}^{*}$. 
On the other hand, at radii sufficiently far
from the SMBH, i.e., at $x=3$ the outflows 
become more supersonic than for models $M_{2}^{*}$ and
$M_{3}^{*}$ as we may see by comparing Figs. 2 and 4. 
These results clearly show that radial
velocities can differ from system to 
system depending on the details of the radiative processes
dominating the source.
\begin{figure}
\begin{center}
\includegraphics[width=8.5cm]{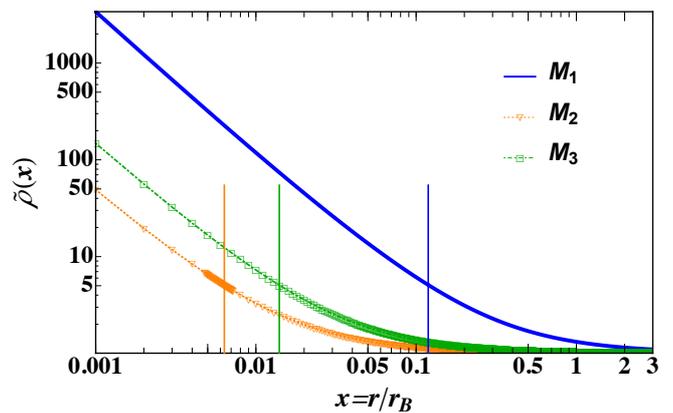}
\end{center}
\caption{Density profiles of the radiative non-isothermal models with $M(\tau)=0$. The vertical
lines mark the position of the sonic radius for each model. As the UV emission becomes stronger,
the accreting gas becomes supersonic at larger radial distances from the SMBH, while the
density becomes higher for $x\lesssim 0.2$. Note that these trends apply only to models $M_{2}$
	and $M_{3}$, while model $M_{1}$, having an arbitrary $\theta _{0}$ and weaker UV emission,
becomes supersonic at a much larger distance from the SMBH and achieves much higher central
densities than models $M_{2}$ and $M_{3}$.}
\end{figure}

For our choice of $\theta_0=\theta _{\rm acc}^{\rm phase}$, Fig. 5 shows the density 
profiles for models
$M_{1}$, $M_{2}$ and $M_{3}$. We see that model $M_{1}$ reaches a density as high as 
$\approx 3000$ at $x=0.001$, which is about 30 times larger than for model $M_{1}^{*}$.
This result implies that when X-ray emission is strong enough, the accretion rate will
also increase by the same factor and the accretion lifetime will decrease for systems with
the same gas reservoirs. In contrast, as the UV emission intensity dominates over the X-ray
emission, the effects of line driving are those of increasing the density close to the
SMBH. In particular, comparing Figs. 3 and 5 we may see that the density at $x=0.001$ is
from 10 to 5 times higher for models $M_{2}^{*}$ and $M_{3}^{*}$ than for models $M_{2}$
and $M_{3}$.  
\begin{figure}
\begin{center}
\includegraphics[width=8.5cm]{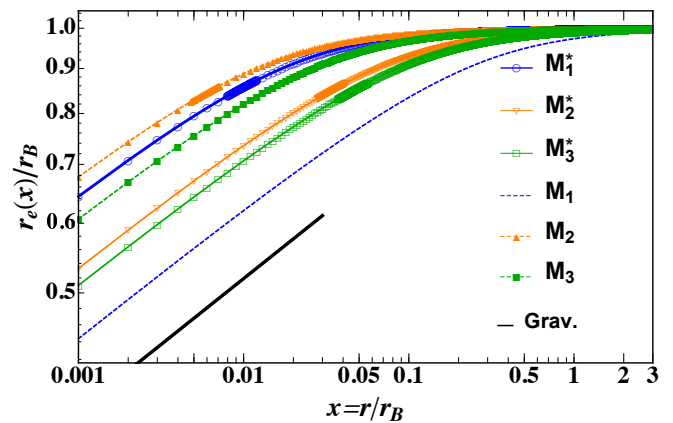}
\end{center}
\caption{Estimated Bondi radius for all models. Models with $M(\tau)\neq 0$ are represented by the
solid lines, while models with $M(\tau)=0$ are depicted by the dashed lines. For comparison, the
black solid line displays the asymptotic behaviour for the pure gravitational case. In the limit
when $x \to \infty$, $r_e\to r_B$.} 
\end{figure}
\begin{figure}
\begin{center}
\includegraphics[width=8.5cm]{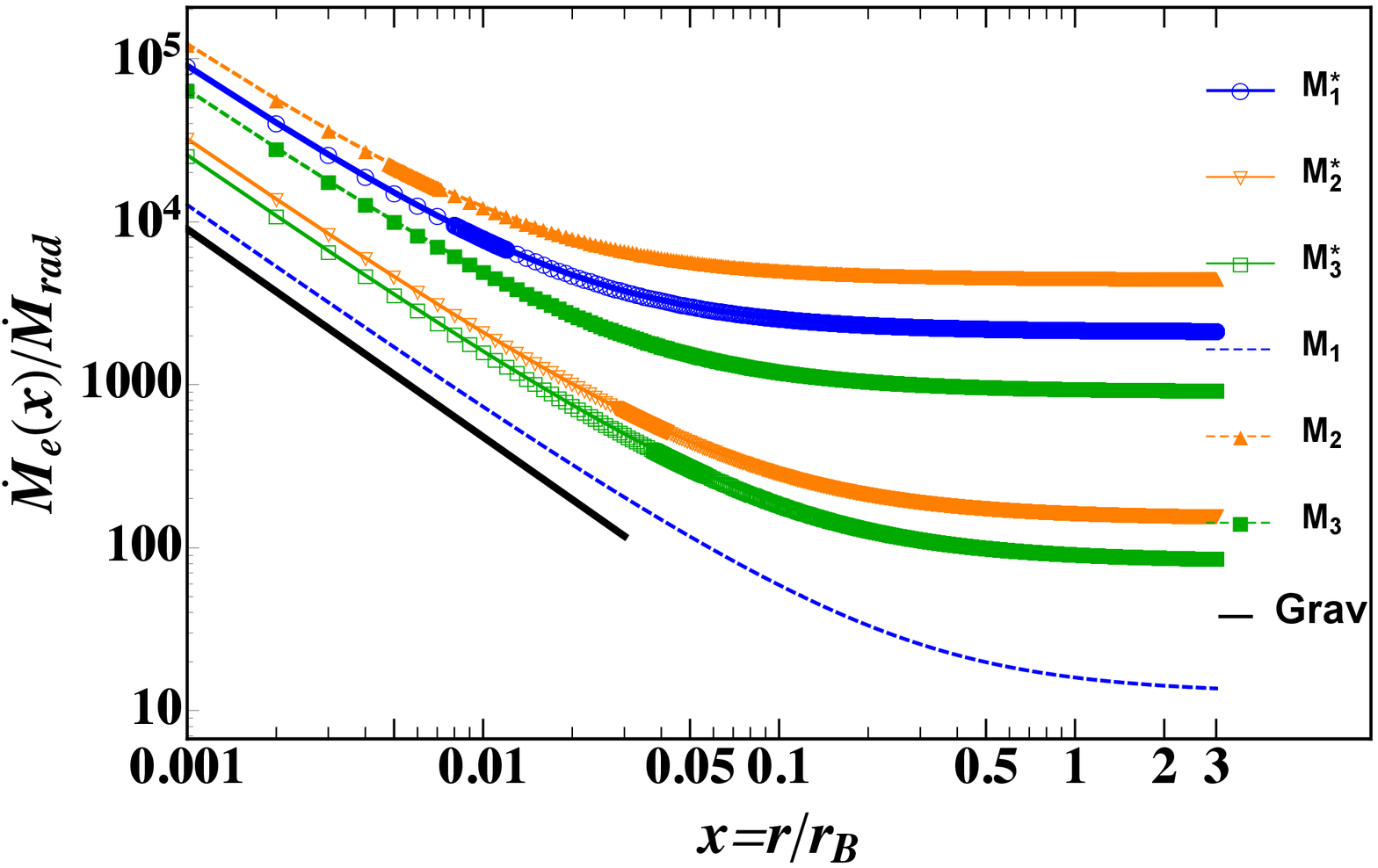}
\end{center}
\caption{Estimated accretion rate for all models. Models with $M(\tau)\neq 0$ are represented by the
solid lines, while models with $M(\tau)=0$ are depicted by the dashed lines. For comparison, the
black solid line displays the asymptotic behaviour for the pure gravitational case.}
\end{figure}

\section{Predicted pure absorption spectral line shape}
\begin{figure}
\begin{center}
\includegraphics[width=8.5cm]{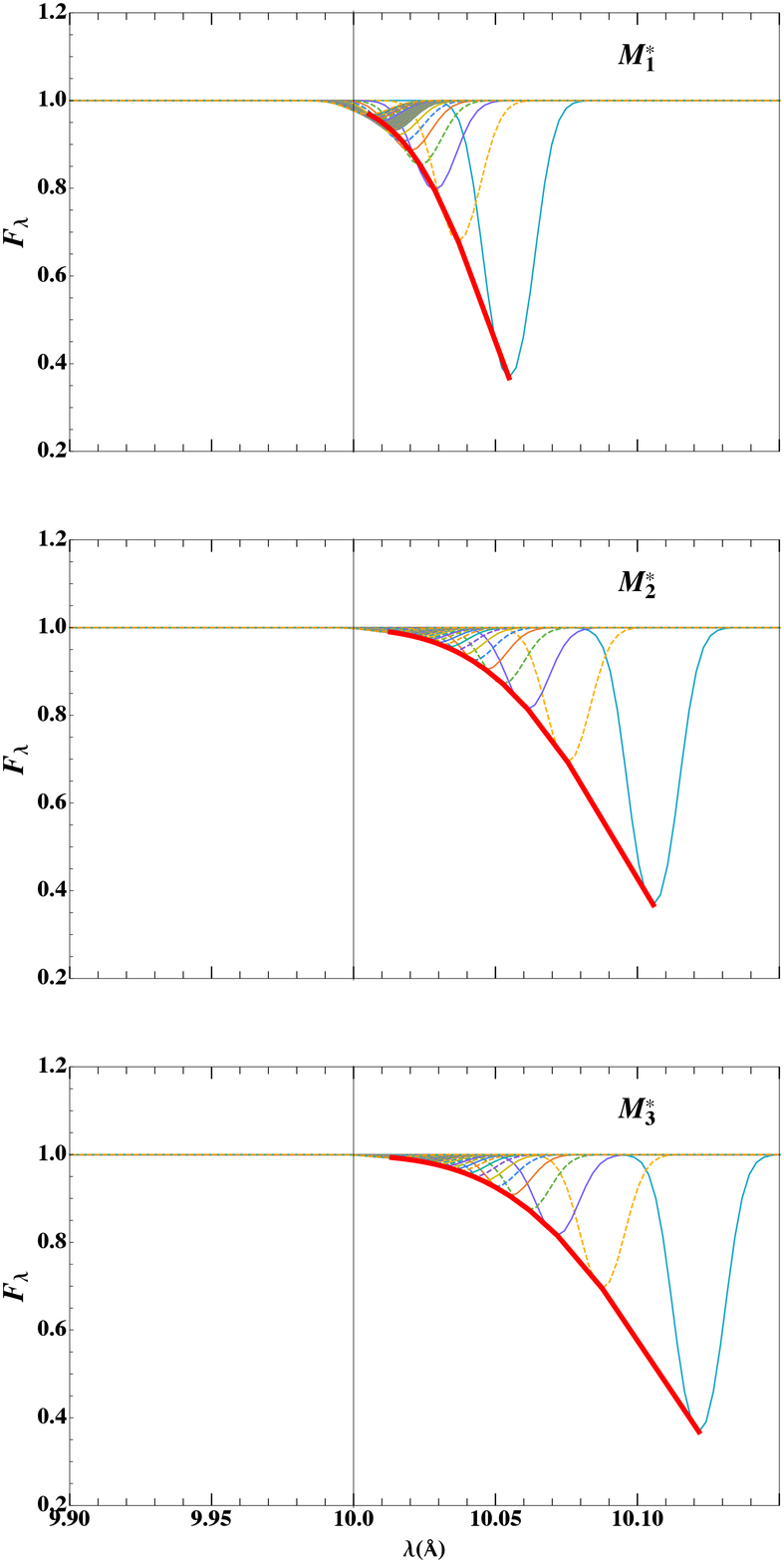}
\end{center}
\caption{Predicted accretion absorption line shape
for models $M_{1}^{*}$, $M_{2}^{*}$ and $M_{3}^{*}$.
In each case, the solid red line shows the profile of a typical
absorption line for an atom falling onto the BH.} 
\label{lineshape}
\end{figure}

A very important prediction of these models is the symmetry
in shape of absorption spectral lines as seen by an
observer far away from the system. To see that we imagine
an atom falling onto the black hole as predicted by model
$M^{*}_{1}$. That atom has a hypothetical transition
which would absorb photons emitted by the inner region
at 10 $\AA$ in the rest-frame ($\lambda _{0}$). 
Then we use the Doppler-shift
formula \citep{rybicki1979a}:
\begin{equation}
w=\frac{w_0}{\gamma L(1-{\rm v}(x)\cos\phi /c)},
\end{equation}
where $w$ is the angular frequency
(measured by the observer) of an emitted photon with
rest-frame frequency $w_0$. As it is used, $\gamma L$
is the Lorentz factor, ${\rm v}(x)$ is the velocity of the
absorbing atom and $\phi$ is the angle between the
streamline and the line-of-sight towards the observer,
which we have set to $\phi=0$.
For falling particles we use ${\rm v}(x)=-\mathcal{M}(x)c_s$,
where $c_s$ is given by equation (7)
\footnote{With $T=10^7$ K, $\gamma=1.1$, and $\mu=0.1$},
for a value of $\approx 345$ km~s$^{-1}$. The simplest absorption
model we could assume for the description of the absorption
spectrum is
\begin{equation}
F_{\lambda}(x)=\exp\left[-A(x)\exp\left(-\frac{(\lambda (x)-\lambda _{0})^2}
{\sigma^2}\right)\right],
\end{equation}
where $\nu _{0}\lambda _{0}=(w_{0}/2\pi)\lambda _{0}=c$ and $\sigma$
is a representative width given by $0.1\% \lambda(x)$.
We set $A(x)=\rho(x)/\rho_{\rm max}$ to model the
intensity of the absorption taken from the density
profile of each model, $\rho(x)=\tilde{\rho}(x)\rho_{\infty}$ and $\rho_{\rm max}=\tilde{\rho}(x=0.001)\rho_{\infty}$. 
As the particle is getting
closer to the BH, the density is increasing and
the final shape will tend to have a deeper deep
shifted towards the red. From Figure
\ref{lineshape}, it is clear that the most
asymmetrical line is the one with $f_{\star}=0.05$
(bottom panel, solid red line),
for which the high-energy radiation is weaker leaving
the particle to reach larger velocities and more
asymmetry. On the other hand, we do not discuss the outflow
(type 2 solution) profile, because the boundary conditions
are not located at the base of the wind as we mentioned earlier.
A deep discussion into the matter of
asymmetrical absorption line shape is beyond the scope
of the present work. An observational treatment
in the Seyfert galaxy NGC 3783 can be found in
\citep{ramirez2005a}. We leave the
detailed description of infalling and outflowing
real atoms and ionized species, like O~{\sc vii},
O~{\sc viii}, Fe~{\sc xxv} for future work, which are of extreme importance
for our understanding of BH science.

\section{Estimated Bondi radius and mass accretion applications}

An important application of the present analysis is to quantify the differences between
the true ($r_{B}$) and estimated ($r_{e}$) Bondi radius as well as the true ($\dot{M}_{\rm rad}$)
and estimated ($\dot{M}_{e}$) accretion rate when the instrumental resolution is limited or
the numerical resolution is inadequate at the sub-parsec scales. Following KCP16 and CP17, these
differences as a function of radial distance from the central source are defined by the
relations
\begin{equation}
\frac{r_{e;\{{\rm acc},{\rm out}\}}(x)}{r_{B}}=\left(\frac{x^2\mathcal{M}_{\{{\rm acc},{\rm out}\}}}
{\chi _{\rm tot}^{\rm rad}|_{\theta=\theta_0}^2\lambda _{\rm cr}}
\right)^{2(\gamma -1)/(\gamma +1)},
\end{equation}
for the radius and
\begin{equation}
\frac{\dot{M}_{e;\{{\rm acc},{\rm out}\}}(x)}{\dot{M}_{\rm rad}}=
\frac{1}{\chi _{\rm tot}^{\rm rad}|_{\theta=\theta_0}^2}\left[\frac{r_{e;\{{\rm acc},{\rm out}\}}(x)}
{r_{B}}\right]^{-(5-3\gamma)/[2(\gamma -1)]},
\end{equation}
for the accretion rate, 
where $\chi _{\rm tot}^{\rm rad}(\theta)$ is the angular-dependent
radiative factor used in this work. 
In Fig. 6 we show the estimated values of the Bondi radius for
all models. For comparison the solid 
black line is the estimated Bondi radius for the
classical Bondi accretion problem. 
At small radii, i.e., at $x=0.001$, model $M_{1}^{*}$ has an
estimated Bondi radius which is 
$\approx 0.64r_{B}$ compared to model $M_{3}^{*}$, which has
$r_{e}/r_{B}\approx 0.51$ when the UV emission from the accretion disk dominates the radiation
field. These values are comparatively lower than those resulting for models $M_{2}$ and $M_{3}$
with no spectral line driving, which have $r_{e}/r_{B}\approx 0.68$ and $\approx 0.6$,
respectively. Moreover, comparing the asymptotic behaviour of $r_{e}/r_{B}$ for the pure
gravitational case (solid black line) given by $r_{e}(x)/r_{B} \sim x^{3(\gamma-1)/2}$ when
$x\to 0^{+}$ (see equation (22) of KCP16), we find differences of $\approx 10\%-40\%$ between the
classical and present generalized Bondi models. In addition, for models $M_{1}$, $M_{2}$ and
$M_{1}^{*}$ the ratio $r_{e}/r_{B}\to 1$ at $x\approx 0.1$, while at larger radii all models
have values of $r_{e}/r_{B}$ close to unity, as shown in Fig. 6 for $x=3$.

The estimated accretion rates for all models are displayed in Fig. 7 as compared with the
asymptotic behaviour of the pure gravitational accretion at small radii given by
$\dot{M}_{e}(x)/\dot{M}_{B}\sim x^{-3(5-3\gamma)/4}$ (see equation (23) of KCP16). We may see
that the radiative effects lead to an overestimation of the accretion rates. In particular at
$x=0.001$, models $M_{1}$ and $M_{2}$ with $M(\tau)=0$ have accretion rates that are from 2 to 5
times larger than models $M_{1}^{*}$ and $M_{2}^{*}$. The differences between these models grow
up to a factor of 10 at $x\approx 3$. The level of overestimation of the accretion rates
compared to the classical Bondi problem is smaller when the effects of line driving are
ignored (in $M_{1}$).
\begin{figure}
\begin{center}
\includegraphics[width=8.5cm]{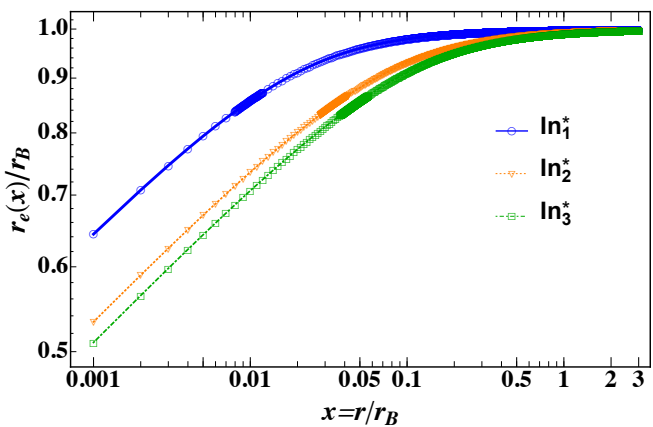}
\end{center}
\caption{Estimated Bondi radius for all models with $M(\tau)\neq 0$. The solid 
line is for the $M_{1}^{*}$, dotted 
line is for the $M_{2}^{*}$, and
dotdashed 
line is for the $M_{3}^{*}$ infalling material.}
\end{figure}
\begin{figure}
\begin{center}
\includegraphics[width=8.5cm]{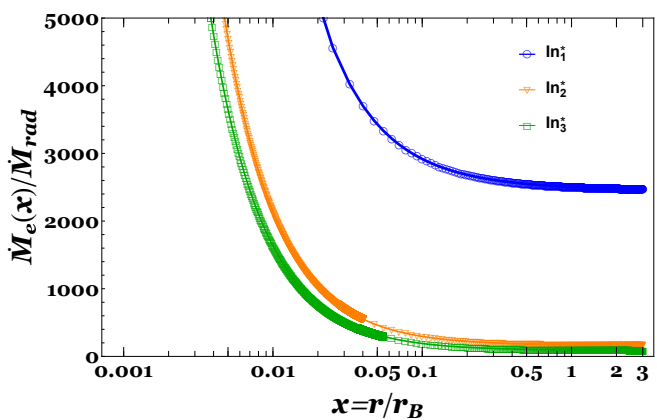}
\end{center}
\caption{Estimated accretion rate for all models with $M(\tau)\neq 0$. The solid 
line is for the $M_{1}^{*}$, dotted 
line is for the $M_{2}^{*}$, and
dotdashed 
line is for the $M_{3}^{*}$ infalling material.
}
\end{figure}

Figure 9 shows the estimated values of the Bondi radius for models $M_{1}^{*}$, $M_{2}^{*}$
and $M_{3}^{*}$ for the inflow solutions (alone). 
For the type 5 solutions the limit
of $r_{e}/r_{B}$ when $x\to\infty$ is no longer one, 
but deviates from it by about 10\%.
This occurs because $r_{B}$ looses its meaning
for $\chi_{\rm tot}^{\rm rad}|_{\theta=\theta_0}\leq 0$,
which is precisely the case here. 
For completeness, Fig. 10 displays the
accretion rates as a function of the radial distance from the SMBH (with y-linear scale). 
The
accretion rates associated with the type 5 
flows are close to those associated with the infalling
material with the differences not being larger than about 3\%.

A further important application of these semi-analytical models is the construction of proper
initial conditions for studying the stability of the Bondi accretion flow \citep{foglizzo05a}.
In most simulations of accretion flows the accretor is assumed to be stationary and small-scale
density and velocity gradients that develop near it, which are the result of non-linear
amplification of numerical noise, may lead to different flow patterns as the resolution is
increased. This noise arises because mutually repulsive pressure forces do not cancel in all
directions simultaneously, giving rise to non-radial velocities whose magnitudes are larger near
the accretor. On the other hand, the use of standard artificial viscosity with a constant coefficient
leads to spurious angular momentum advection in the presence of vorticity. However, the intrinsic
noise generated due to numerical effects is not sufficient to make the flow unstable at parsec
scales, while it may become prominent in the proximity of the accretor at sub-parsec scales.
On the other hand, the strength and type of the instability may also depend on the size of the
accretor \citep{Blondin2012}, the Mach number and the polytropic index $\gamma$. Understanding
how these parameters can influence noise amplification at sub-parsec scales will shed light on the
instability production mechanisms and the accretion rate for long-stage accretion systems.

\section{Summary and conclusions}

We have presented the 
classical Bondi accretion theory for the case of
non-isothermal accretion processes onto a 
supermassive black hole (SMBH), including the
effects of X-ray heating and the 
radiation force due to electron scattering and spectral
lines. The radiation field is modelled following the recipe of 
P07,
where an optically thick, geometrically thin, 
standard accretion disk is considered as a
source of UV photons and a spherical central object as a source of X-ray emission. 
A semi-analytical solution for the radiative non-isothermal accretion onto a SMBH at the
centre of galaxies is obtained using a procedure similar to that developed by \cite{korol16a}
(KCP16). A novel feature of the present analysis is the angular dependence of the UV emission
from the accretion disk, while the X-ray/central object radiation is assumed to be isotropic.
The influence of both types of radiation is evaluated for different flux fractions of the
X-ray/central object radiation ($f_{\star}$) and the UV emission from the disk
($f_{\rm disk}$) with and without the effects of spectral line driving 
for an incident angle $\theta_0 =\theta _{\rm acc}^{\rm phase}$.

The main conclusions can be summarized as follows:
\begin{enumerate}

\item The ratio between the X-ray and the UV flux increases with increasing $\theta _{0}$.
For $\theta _{0}=\pi /2$, the radiation flux from the accretion disk vanishes and so only the
X-ray emission contributes to the radiation flux.

\item When the radiation force due to spectral lines does not contribute to the heating,
pure gravitational accretion onto the SMBH occurs when the X-ray emission intensity is
the strongest and has the same fraction of participation as the UV emission (i.e.,
$f_{\star}=f_{\rm disk}=0.5$). As long as the UV emission dominates over the X-ray
heating, a transition from pure accretion to mathematical, but non trivial physical, type 5 
flows occurs.

\item When the radiation force due to spectral lines is taken into account, a 
transition from pure accretion to outflows always takes place independently of the intensity
of the X-ray emission. However, as the UV emission dominates over the X-ray heating,
the angle of incidence for which the pure accretion/outflow transition occurs and the
radial distance from the SMBH for which the inflow and the outflow becomes supersonic
both increase.

\item As the UV emission becomes stronger, the gas density close to the centre becomes
higher. The same trend also applies to the cases where line driving is not considered,
except when the X-ray luminosity is strong enough to suppress the outflow. In this case,
only pure accretion takes place and the central density achieves much higher values than
the other models.

\item For our radiative, non-isothermal Bondi 
accretion model, we also provide
the exact formula for the Bondi radius bias $r_{e}/r_{B}$ as a function of the radial
distance from the central accretor, the Mach number, the critical accretion parameter and
the total radiation luminosity. The exact formula for the mass accretion bias
$\dot M_{e}/\dot M_{B}$ is also given in terms of $r_{e}/r_{B}$.

\item The estimated values of the Bondi radius $r_{e}$ are between $\approx 0.2$ and
$\approx 0.68$ times the classical Bondi radius $r_{B}$ close to the accretor, while
$r_{e}\to r_{B}$ at large distances from the accretor ($r\gtrsim 3r_{B}$).

\item The radiative effects produce an overestimation of the estimated accretion rates
compared to the classical Bondi accretion rates. However, the level of overestimation
is smaller when the effects of line driving are ignored for the $M_1$ models.

\item Under the effects of line driving, the limit of $r_{e}/r_{B}$ 
(for $\chi_{\rm tot}^{\rm rad}|_{\theta=\theta_0}\leq 0$) at large distances
from the central accretor is no longer one, but deviates from unity by about 10\%. On
the other hand, the accretion rates associated 
with $(\chi_{\rm tot}^{\rm rad}|_{\theta=\theta_0}\leq 0)$--flows
are close to those
associated with the accreting material with differences $\lesssim 3$\%.

\item The prediction of asymmetric pure absorption lines
as seen in the line-of-sight of a distant observer,
serves as platform to construct more complex scenarios
by adding emission components and/or including ionization
structure to the interacting gas. This is becoming
imprint in the light of the new generation of telescopes
that are able of imaging the centres of galaxies as well as the spectroscopy
of the centres of galaxies and the new BH science.

\item {\bf We have computed the radiative factor for a $\xi_{\rm 10pc}=\xi(r={\rm 10~pc})$
without temperature corrections although the hydrodynamical variables were taken to be
temperature dependent. A future work in this line will consider the temperature and 
ionization dependence.}

\end{enumerate}

We conclude by emphasizing that the present results are useful to model proper initial
conditions for time-dependent simulations of accretion flows onto massive black holes
at the centre of galaxies. A further important application of the present semi-analytical
solutions concerns the stability of the Bondi accretion flow at sufficiently close
distances from the accretor, where the stability of the flow is expected to be affected by
the growth of small density and velocity fluctuations. As a further step in this line of
research, we plan to include the additional effects of the gravitational potential of the
host galaxy \citep{korol16a} for the cases of Hernquist and Jaffe galaxy models \citep{ciotti17a},
which are applicable to early-type galaxies.

\section{Acknowledgments}

JMRV thanks D. Proga for in-deep discussion 
into the X type solutions of the Bondi problem.
{\bf J. K. acknowledges financial support by 
the Consejo Nacional de Ciencia y Tecnolog\'{\i}a 
(CONACyT), Mexico, under grant 283151. 
We are really indebted with the anonymous 
referee who has raised a number of suggestions
that have improved and clarified the content of the manuscript.}
{\sc impetus} is a collaboration project 
between the {\sc abacus}-Centro de Matem\'atica
Aplicada y C\'omputo de Alto 
Rendimiento of Cinvestav-IPN, the School of Physical Sciences
and Nanotechnology, Yachay Tech University, 
and the \'Area de F\'{\i}sica de Procesos
Irreversibles of the Departamento de 
Ciencias B\'asicas of the Universidad Aut\'onoma
Metropolitana--Azcapotzalco (UAM-A) 
aimed at the SPH modelling of astrophysical flows. 
This work has been partially supported by 
Yachay Tech under the project for the use of the 
supercomputer {\sc quinde i} and by 
UAM-A through internal funds. We also thank the support 
from Cinvestav-Abacus where part of this work was done.

\bibliography{Final_BIBTEX.bib}

\end{document}